\documentclass[prc,aps,superscriptaddress,twocolumn]{revtex4}
\usepackage{graphicx}
\usepackage{color}
\begin{document}
\title{Nuclear and neutron matter equations of state from high-quality potentials up     
to \\ fifth order of the chiral expansion} 

\author{F. Sammarruca}
\affiliation{Department of Physics, University of Idaho, Moscow, ID 83844, USA}

\author{L. E. Marcucci}
\affiliation{Dipartimento di Fisica ``Enrico Fermi'', Universit\`a
di Pisa, Largo Bruno Pontecorvo 3 - I-56127 Pisa, Italy}
\affiliation{Istituto Nazionale di Fisica Nucleare, Sezione di Pisa,\\
Largo Bruno Pontecorvo 3 - I-56127 Pisa, Italy}

\author{L. Coraggio}
\affiliation{Istituto Nazionale di Fisica Nucleare, Sezione di Napoli\\
Complesso Universitario di Monte S. Angelo, Via Cintia - I-80126 Napoli, Italy}

\author{J. W. Holt}
\affiliation{Cyclotron Institute and Department of Physics and Astronomy, Texas A \& M University,                 
College Station, TX 77843 USA}

\author{N. Itaco}
\affiliation{Istituto Nazionale di Fisica Nucleare, Sezione di Napoli\\
Complesso Universitario di Monte  S. Angelo, Via Cintia - I-80126 Napoli, Italy}
\affiliation{Dipartimento di Matematica e Fisica, Universit\`a della
  Campania ``Luigi Vanvitelli'', \\
Viale Lincoln 5, I-81100, Caserta, Italy}

\author{R. Machleidt}
\affiliation{Department of Physics, University of Idaho, Moscow, ID 83844, USA}

\begin{abstract}
We present predictions for the equation of state of symmetric nuclear
and pure neutron matter based on recent high-quality nucleon-nucleon potentials 
from leading order to fifth order in the chiral expansion. We include as well the next-to-next-to-leading
order (N$^2$LO) chiral three-nucleon force 
whose low-energy constants $c_D$ and $c_E$ are fitted to the binding 
energies of $^3$H and $^3$He as well as the $\beta$-decay lifetime of $^3$H.
The ground state energy per particle is computed in the particle-particle ladder approximation up 
to a few times saturation density. Due to the soft character of the interactions, uncertainties due to
the convergence in many-body perturbation theory are small. We find that nuclear matter saturation 
is reproduced quantitatively at N$^3$LO and N$^4$LO, and therefore we encourage the application 
of these interactions in finite nuclei, where the description of ground-state energies and charge radii 
of medium-mass nuclei may be improved.
\end{abstract}
\maketitle 
        
\section{Introduction} 
\label{Intro} 

The properties of neutron-rich matter are important for addressing a number of open questions in 
nuclear physics and nuclear astrophysics, including the location of neutron drip lines, the thickness 
of neutron skins, and the structure of neutron stars. These questions have in common
a strong sensitivity to the nuclear equation of state (EoS), namely the energy per 
particle as a function of density and composition (set by the isospin asymmetry $\delta_{np} = 
(\rho_n-\rho_p)/(\rho_n+\rho_p)$, where $\rho_n$ and $\rho_p$ are the neutron and proton
number densities, respectively). The symmetry energy, which is defined as the difference in the
energy per particle of pure neutron matter and symmetric nuclear matter at a given density, determines
to a good approximation also the energy per particle of homogeneous nuclear matter with arbitrary 
isospin asymmetry. The 
symmetry energy and its density dependence are therefore a key focus of contemporary theoretical 
and experimental investigations, and much effort has been devoted to finding correlations between
nuclear observables and this property of infinite matter.

Constructing the EoS microscopically from state-of-the-art few-body interactions gives fundamental
insight into effective nuclear forces in the medium. 
High-precision meson-theoretic interactions~\cite{Mac01,Sto94,WSS95} are still frequently employed in contemporary calculations of nuclear matter, structure and reactions. However, in this framework 
three-nucleon forces (3NFs), or more generally $A$-nucleon forces with $A>2$, have only a loose 
connection with the associated two-nucleon force (2NF)~\cite{Mac17}, and there exists no clear 
scheme to quantify and control the theoretical uncertainties. Chiral effective field theory (EFT)~\cite{ME11,EHM09,MS16}, on the other hand, provides a more systematic approach to construct 
nuclear many-body forces, which emerge on an equal footing~\cite{Wei92} with two-body forces, and 
to assess theoretical uncertainties through a systematic expansion controlled by the
``power counting"~\cite{Wei90}. Furthermore, chiral EFT maintains consistency with the symmetries and
symmetry breaking pattern of the underlying fundamental theory of strong interactions, quantum 
chromodynamics (QCD). 

For the reasons described above, chiral EFT has evolved into the authoritative approach for 
developing nuclear forces, and modern applications have focused on few-nucleon 
reactions~\cite{Epe02,NRQ10,Viv13,Gol14,Kal12,Nav16}, the structure of light- and medium-mass nuclei~\cite{Coraggio07,Coraggio10,Coraggio12,Hag12a,Hag12b,BNV13,Gez13,Her13,Hag14a,Som14,Heb15,Hag16,Car15,Her16,Hol17,Sim17,Mor17},
infinite matter at zero temperature~\cite{HS10,Heb11,Baa13,Hag14b,Cor13,Cor14,Sam15,Dri16,Tew16,MS16,Hol17}
and finite temperature~\cite{Wel14,Wel15}, and nuclear dynamics and response functions \cite{Bac09,Bar14,Rap15,Bur16,Hol16,Bir17,Rot17}.
Although satisfactory predictions have been obtained in many cases, specific problems persist.
These include the description of bulk properties of medium-mass nuclei, which typically
exhibit charge radii that are too small~\cite{Lap16} and binding 
energies that are highly sensitive to the choice of nuclear force and often turn out to be 
too large~\cite{Bin14}. This has led some groups to fit the low-energy constants that parametrize
unresolved short-distance physics in chiral nuclear forces directly to the properties of 
medium-mass nuclei~\cite{Eks15} and, indeed, better predictions for other isotopes are then 
obtained. However, one would prefer a genuine microscopic approach in which the 2NF is fixed 
by two-nucleon data and the 3NF by three-nucleon data, with no further fine tuning. Applications to 
systems with $A>3$ would then be true predictions, though possibly with large uncertainties.

Two recent studies~\cite{Sim17,Mor17} provide indications for how the overbinding problem may 
be overcome. In these studies, a rather soft nucleon-nucleon ($NN$) potential (due to renormalization 
group evolution) together with 3NFs refitted to the $^3$H binding energy and the $^4$He charge radius 
were used to calculate the ground-state properties of closed shell nuclei ranging from $^4$He to the light 
Tin isotopes~\cite{Sim17,Mor17}. The ground-state energies were reproduced accurately, while the radii 
came out slightly too small, but not dramatically different from experiment. These features can be
linked to the good nuclear matter saturation properties of the employed 2NF + 3NF 
combination~\cite{Heb11}. In the above example, the 2NF was soft and alone would lead to 
substantial overbinding in nuclear matter, but the addition of a repulsive 3NF contribution leads to 
a much better description of the nuclear matter saturation point~\cite{Heb11}. On a historical note, 
the first quantitative explanation for nuclear matter saturation was achieved in this way within the
framework of Dirac-Brueckner-Hartree-Fock theory~\cite{BM90,AS03,Sam08,Sam12,MSM17}.
Alternatively, one may start from a relatively repulsive 2NF and then add an attractive, density-dependent 
3NF contribution, such as the combination of the Argonne $v_{18}$ (AV18) 2NF~\cite{WSS95} plus the 
Urbana IX 3NF~\cite{Pud95}. However, the nuclear matter saturation energy and density cannot 
be simultaneously reproduced by this combination~\cite{APR98} and medium-mass nuclei are severely underbound~\cite{Lon17}. Similar problems occur when the AV18 2NF is combined with the Illinois-7 
3NF~\cite{Pie08,Lon17}.

In Ref.~\cite{EMN}, high-quality soft chiral $NN$ potentials from leading order to 
fifth order in the chiral expansion were constructed. These interactions are more consistent than those 
constructed earlier~\cite{EM03,chinn5,ME11}, in the sense that the same power counting scheme and 
cutoff procedures are used at all orders. For these potentials, the very accurate $\pi N$ low-energy 
constants (LECs) determined in the Roy-Steiner analysis of Ref.~\cite{Hofe+} are applied. The 
uncertainties associated with these LECs are so small that variations within the errors have negligible 
impact on the construction of the potentials. That the potentials are soft and rather perturbative
has been demonstrated in the investigations of Refs.~\cite{Hop17,DHS17}.
In the present work, we extend a subset of the chiral nuclear forces in Ref.\ \cite{EMN} to include 
the consistent N$^2$LO three-body force with LECs fitted to the binding energies of $A=3$ nuclei
and the triton $\beta$-decay lifetime. We then present predictions for the EoS of symmetric nuclear matter 
(SNM) and pure neutron matter (PNM) based on this family of chiral potentials. We explore the 
resulting saturation properties, the convergence pattern in the chiral expansion, the impact of the 3NFs 
at each order, individual contributions of the 3NF, regulator
dependence, and the behavior of these
  potentials with respect to a perturbative expansion of the ground
  state energy per nucleon.

The manuscript is organized as follows: in Secs.~\ref{II} and~\ref{III} we briefly
summarize the main features of the 2NFs and 3NFs employed in this work.
The reader is referred to Ref.~\cite{EMN} for a complete and
detailed description of the 2NF. In Sec.~\ref{IV} we present our results
for SNM and PNM. Finally, our conclusions are summarized in Sec.~\ref{Concl}. 
     
\section{The two-nucleon forces}  
\label{II} 

The $NN$ potentials employed in this work span five orders in the chiral EFT expansion, 
from leading order (LO) to fifth order (N$^4$LO). The same power counting scheme and 
regularization procedures are applied through all orders, making this set of interactions 
more consistent than previous ones. 
Another novel and important aspect in the construction of these new potentials is the fact 
that the long-range part of the interaction is fixed by the $\pi N$ LECs as determined in the 
recent and very accurate analysis of Ref.~\cite{Hofe+}. In fact, for all practical purposes, 
errors in the $\pi N$ LECs are no longer an issue with regard to uncertainty quantification.
Furthemore, at the fifth (and highest) order, the $NN$ data below pion production threshold 
are reproduced with excellent precision ($\chi ^2$/datum = 1.15).

Iteration of the potential in the Lippmann-Schwinger equation requires                         
cutting off high-momentum components, consistent with the fact that chiral perturbation theory 
amounts to building a low-momentum expansion.          
This is accomplished through the application of a regulator function for which the non-local form
is chosen:
\begin{equation}
f(p',p) = \exp[-(p'/\Lambda)^{2n} - (p/\Lambda)^{2n}] \,,
\label{reg}
\end{equation}
where $p' \equiv |{\vec p}\,'|$ and $p \equiv |\vec p \, |$ denote the final and initial nucleon momenta in the center-of-mass system, respectively.
For the present applications in nuclear and neutron matter, we will limit ourselves to values of the 
cutoff parameter $\Lambda$ smaller than or equal to 500 MeV, as those have been associated 
with the onset of favorable perturbative properties. 
The soft nature of the potentials has been confirmed by the Weinberg eigenvalue analysis of 
Ref.~\cite{Hop17} and in the context of the perturbative calculations of infinite matter of 
Ref.~\cite{DHS17}.

\section{The three-nucleon forces} 
\label{III} 

\begin{figure}[t]
\centering
\includegraphics[scale=0.65]{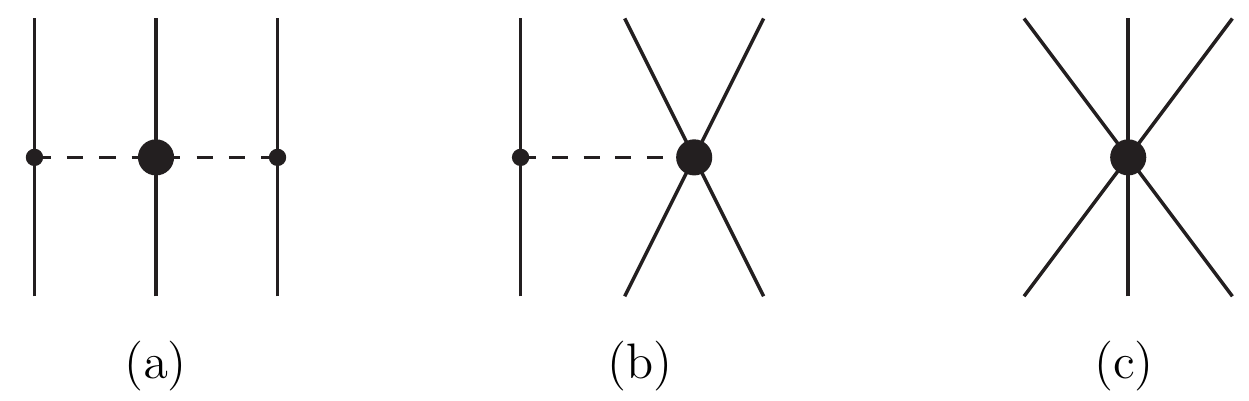}
\caption{The 3NF at N$^2$LO with (a) the 2PE, (b) the 1PE, and (c) the contact diagrams.} 
\label{fig1}
\end{figure}

Three-nucleon forces make their first appearance at the third order of the chiral expansion 
(N$^2$LO). At this order, the 3NF consists of three contributions~\cite{Epe02}:
the long-range two-pion-exchange (2PE) term, 
the medium-range one-pion exchange (1PE) diagram, and a short-range contact term. 
The corresponding diagrams are shown in Fig.~\ref{fig1}.
We apply these 3NFs by way of the density-dependent effective two-nucleon interactions derived in 
Refs.~\cite{holt09,holt10}. They are expressed in terms of the well-known 
non-relativistic two-body nuclear force operators and can be conveniently 
incorporated in the usual $NN$ partial wave formalism and the particle-particle ladder approximation
for computing the EoS.                
The effective density-dependent two-nucleon interactions consist of six one-loop topologies. Three of 
them are generated from the 2PE graph of the chiral 3NF, Fig.~\ref{fig1}(a), and depend on the LECs
$c_{1,3,4}$, which are already present in the 2PE part of the $NN$ interaction. 
Two one-loop diagrams are generated from the 1PE diagram, Fig.~\ref{fig1}(b), and depend on 
the low-energy constant $c_D$. Finally, there is the one-loop diagram that involves the 3NF contact 
diagram, Fig.~\ref{fig1}(c), with LEC $c_E$.
Note that, in pure neutron matter, the contributions proportional to the LECs $c_4,c_D$, and 
$c_E$ vanish~\cite{HS10}. In recent nuclear matter calculations \cite{Dri16,DHS17}, progress has 
been made toward including N$^3$LO three-body interactions in the two-body normal-ordering 
approximation as well as including the residual three-body normal-ordered force.

\begin{figure}[t]
\includegraphics[width=8.1cm]{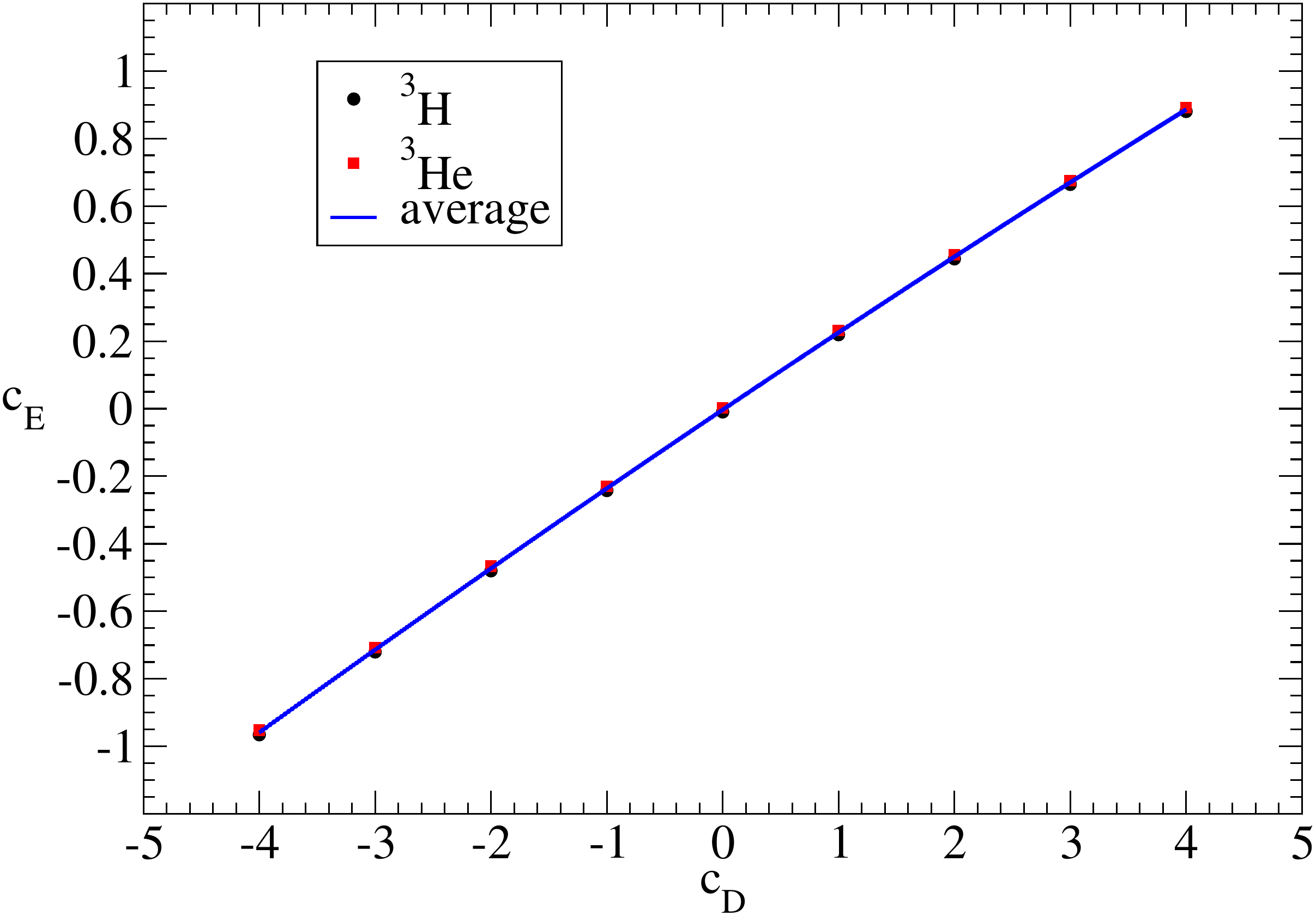}
\caption{Trajectory of $c_D$ and $c_E$ obtained by fitting to the $^3$H or $^3$He 
binding energies using the N$^4$LO $NN$ potential and the N$^2$LO 3NF 
with $\Lambda = 450$ MeV. The 
average trajectory is also shown.}
\label{fig:cd-ce1}
\end{figure}

We fix the LECs $c_D$ and $c_E$ within the three-nucleon sector. Specifically, 
we constrain them to reproduce the $A=3$ binding energies and the Gamow-Teller
(GT) matrix element of tritium $\beta$-decay, following a well
established procedure~\cite{Gar06,Gaz09,Mar12}.
A few comments are here in order.
(i) The regulator function used in the derivation of the
3NF is that of Ref.~\cite{Nav07},
i.e.\
\begin{equation}
  f(q)=\exp[(-q/\Lambda)^4]\ ,
  \label{eq:reg_3NF}
\end{equation}
where $q=| \vec p~'-\vec p \, |$ is the momentum transfer. With this choice, the 3NF is local
in coordinate space, making the construction of the $A=3$
wave functions less involved~\cite{Kie08}.
(ii) The relation which in
Refs.~\cite{Gaz09,Mar12} was used to relate the LEC $c_D$ with $d_R$,
the LEC entering the axial current at N$^2$LO, has been recently
revisited~\cite{Sch_PC,Mar12}. While in Refs.~\cite{Gaz09,Mar12}
it was
\begin{equation}
  d_R=\frac{m}{g_A\Lambda_\chi} c_D +\frac{m}{3} (c_3+2c_4) +\frac{1}{6} \ ,
  \label{eq:dr_old}
\end{equation}
$m$ being the nucleon mass, $g_A$ the single-nucleon axial coupling
constant (here $g_A=$1.2723~\cite{Bar16}),
$\Lambda_\chi$ the chiral-symmetry-breaking scale
(here $\Lambda_\chi=$1 GeV), in Ref.~\cite{Sch_PC} it was shown that
the correct relation is
\begin{equation}
  d_R=-\frac{m}{4g_A\Lambda_\chi} c_D +\frac{m}{3} (c_3+2c_4) +\frac{1}{6} \ .
  \label{eq:dr_new}
\end{equation}
This has been confirmed also in Ref.~\cite{Kre17}. We note that       
the muon capture rates calculated in Ref.~\cite{Mar12} 
have remained unchanged. 
This is not surprising, since the LECs $c_D$ and $c_E$
originally used
in the axial current of the
muon capture (as well as proton weak capture of Ref.~\cite{Mar13})
lead to the same enhancement as in Ref.~\cite{Mar12}
needed to reproduce the tritium GT matrix element.

\begin{figure}[t]
\includegraphics[width=8.cm]{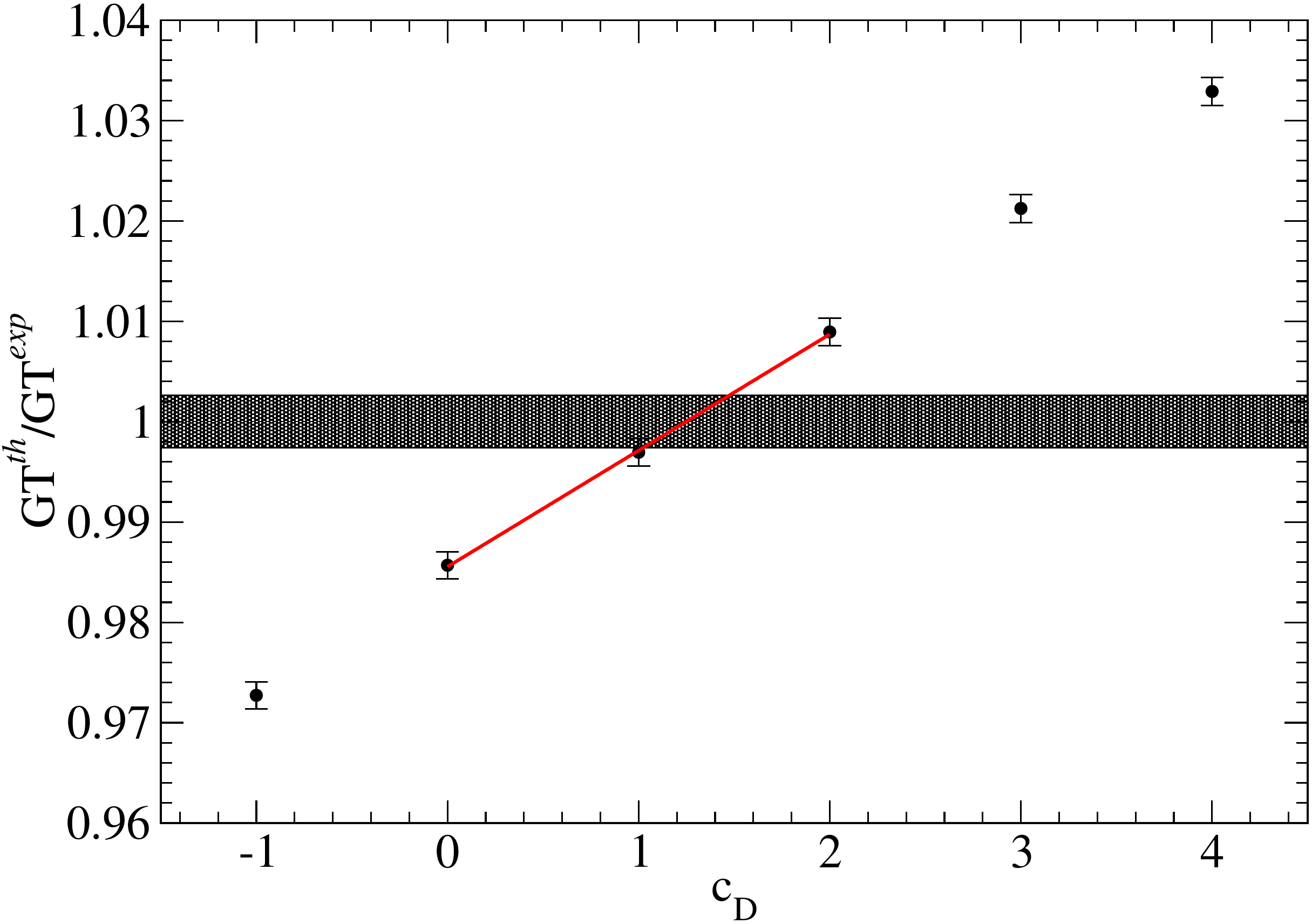}
\caption{Ratio GT$^{th}$/GT$^{exp}$ as a function of $c_D$ using the N$^4$LO $NN$ 
potential and the N$^2$LO 3NF with $\Lambda = 450$ MeV. 
The circles are the calculated values, while the 
solid (red) line is a linear fit in the region where the experimental band is crossed.}
\label{fig:cd-ce2}
\end{figure}

\begin{table*}[t]
\caption{
Values of the LECs $c_{1,3,4}$, $c_D$, and $c_E$ 
for different orders in the chiral EFT expansion and different values of the momentum-space
cutoff $\Lambda$.
The LECs $c_{1,3,4}$ are given in units of GeV$^{-1}$, while
$c_D$ and $c_E$ are dimensionless. The numbers in parentheses
indicate the error arising from the fitting procedure.
In addition, we also show the value for the exponent $n$ that 
appears in the regulator function of Eq.~(\ref{reg}).}
\label{tab1}
\begin{tabular*}{\textwidth}{@{\extracolsep{\fill}}cccccccc}
\hline
\hline
  & $\Lambda$ (MeV) & $n$& $c_1$ & $c_3$ & $c_4$ & $c_D$ & $c_E$ \\
\hline    
\hline
N$^2$LO & 450 & 2& --0.74 & --3.61 & 2.44  &  0.935(0.215) &   0.12(0.04)  \\
        & 500 & 2& --0.74 & --3.61 & 2.44  &  0.495(0.195) & --0.07(0.04) \\
\hline 
N$^3$LO & 450 & 2& --1.07 & --5.32 & 3.56  &  0.675(0.205) &   0.31(0.05)  \\
        & 500 & 2& --1.07 & --5.32 & 3.56  &--0.945(0.215) & --0.68(0.04) \\
\hline 
N$^4$LO & 450 & 2& --1.10 & --5.54 & 4.17  &  1.245(0.225) &   0.28(0.05)  \\
        & 500 & 2& --1.10 & --5.54 & 4.17  &--0.670(0.230) & --0.83(0.03)  \\
\hline
\hline
\end{tabular*}
\end{table*}

\begin{table*}
  \caption{Same as Table~\ref{tab1}, 
    but including the 2PE 3NF 
at fourth and fifth order, respectively. 
(The N$^2$LO numbers are the same as in Table~\ref{tab1}.)
}
\label{tab2}
\centering
\begin{tabular*}{\textwidth}{@{\extracolsep{\fill}}cccccccc}
\hline
\hline
   & $\Lambda$ (MeV) & $n$ & $c_1$ & $c_3$ & $c_4$ & $c_D$ & $c_E$ \\
\hline     
\hline
N$^2$LO & 450 & 2 & --0.74 & --3.61 & 2.44 &  0.935(0.215) &  0.12(0.04) \\
      & 500 & 2 & --0.74 & --3.61 & 2.44 &  0.495(0.195) &--0.07(0.04) \\
\hline 
N$^3$LO & 450 & 2 & --1.20 & --4.43 & 2.67 &  0.670(0.210) &  0.41(0.05) \\
        & 500 & 2 & --1.20 & --4.43 & 2.67 &--0.750(0.210) &--0.41(0.04) \\
\hline
N$^4$LO & 450 & 2 & --0.73 & --3.38 & 1.69 &  0.560(0.220) &  0.46(0.05) \\
        & 500 & 2 & --0.73 & --3.38 & 1.69 &--0.745(0.225) &--0.15(0.04) \\
\hline
\hline
\end{tabular*}
\end{table*}

For the sake of completeness, we recall the various steps for the
adopted fitting procedure, which are: (i) determination of a $c_D-c_E$
trajectory, obtained by reproducing the $^3$H and $^3$He binding
energies. Typically, the $^3$H and $^3$He trajectories
are indistinguishable, and the average can be safely
used, leading to the $A=3$ binding energies within $\sim 10$ keV
of the experimental ones. (ii) For each set of $c_D-c_E$ values,
we have calculated the GT matrix element of tritium $\beta$-decay and fitted to
its experimental value, taken as in Ref.~\cite{Bar16} to be
$0.9511\pm 0.0013$. We have then found a range of $c_D$ values,
for which the theoretical and experimental GT values, GT$^{th}$
and GT$^{exp}$ respectively, coincide (note that
we have conservatively doubled the experimental error on GT$^{exp}$).
The corresponding values for $c_E$ are obtained from the
$c_D-c_E$ trajectory mentioned above. In Table~\ref{tab1}
we list the central values of the LECs $c_D$ and $c_E$ used in this
work, with the error in parentheses arising from the fitting procedure.
Obviously, these values have been obtaining using Eq.~(\ref{eq:dr_new}).
As we can see by inspection of the table, the allowed range for
$c_D$ is quite large, while the corresponding one for $c_E$ is 
much smaller. Since the $c_D$ contribution to the energy per
particle in SNM is very small (see below), we do not include this uncertainty
in the present calculations.
As an example, we show in Fig.~\ref{fig:cd-ce1} the $c_D-c_E$
trajectory and in Fig.~\ref{fig:cd-ce2} the ratio GT$^{th}$/GT$^{exp}$ as a function of $c_D$
for the N$^4$LO $NN$ potential and the N$^2$LO 3NF
with $\Lambda=450$ MeV.

The complete 3NF beyond N$^2$LO is very complex and often neglected in nuclear structure
studies, but progress toward the inclusion of the subleading 3NF at N$^3$LO is underway 
\cite{Tew13,Dri16,DHS17,Heb15a}.
There is one important component of the 3NF where complete
calculations up to N$^4$LO are possible: the 2PE 3NF. In Ref.~\cite{KGE12} it was
shown that the 2PE 3NF has essentially the same analytical structure at N$^2$LO, 
N$^3$LO, and N$^4$LO. Thus, one can add the three orders of 3NF contributions and 
parametrize the result in terms of effective LECs. 

In the N$^4$LO rows of Table~\ref{tab2} we give 
the effective LECs $c_{1,3,4}$ obtained in Ref.~\cite{KGE12}.
Concerning the 2PE 3NF at N$^3$LO, Eq.~(2.8) of Ref.~\cite{Ber08} provides the corrections 
to the $c_i$ but there is an error in the numerical values given below this equation.
While $\delta c_1 = -0.13$ GeV$^{-1}$ is correct, the correct values for $\delta c_3$ and
$\delta c_4$ are $\delta c_3 = -\delta c_4 = 0.89$ GeV$^{-1}$.
When these corrections are applied, the values given in the N$^3$LO rows  of Table~\ref{tab2} 
emerge. By using the ${c}_i$ of Table~\ref{tab2} in the mathematical expression of the N$^2$LO 3NF, 
one can include the 2PE parts of the 3NF up to N$^3$LO and up to N$^4$LO in a simple way. 
Consequently, the LECs $c_D$ and $c_E$ are fitted
again with the same procedure outlined above. Their values are also
listed in Table~\ref{tab2}. They are clearly different from those
listed in Table~\ref{tab1} but of the same order and with the same sign.
Among all possible 3NF contributions, the 2PE 3NF is historically
the first calculated~\cite{FM57}. The prescriptions given above allow us to incorporate 
this very important 3NF up to the highest orders considered in this paper.

\section{Calculations of the equation of state} 
\label{IV} 

We perform microscopic calculations of nuclear and
neutron matter with the interactions described above.
We compute the EoS using the nonperturbative particle-particle
ladder approximation, which generates the leading-order contributions 
in the traditional hole-line expansion. In addition we compute 
the single-particle spectrum for the intermediate-state energies also in the 
particle-particle ladder approximation, keeping only the real part.

\subsection{Nuclear matter: predictions and discussion} 
\label{IVa} 

\begin{figure}[t]
\centering
\includegraphics[width=6.5cm]{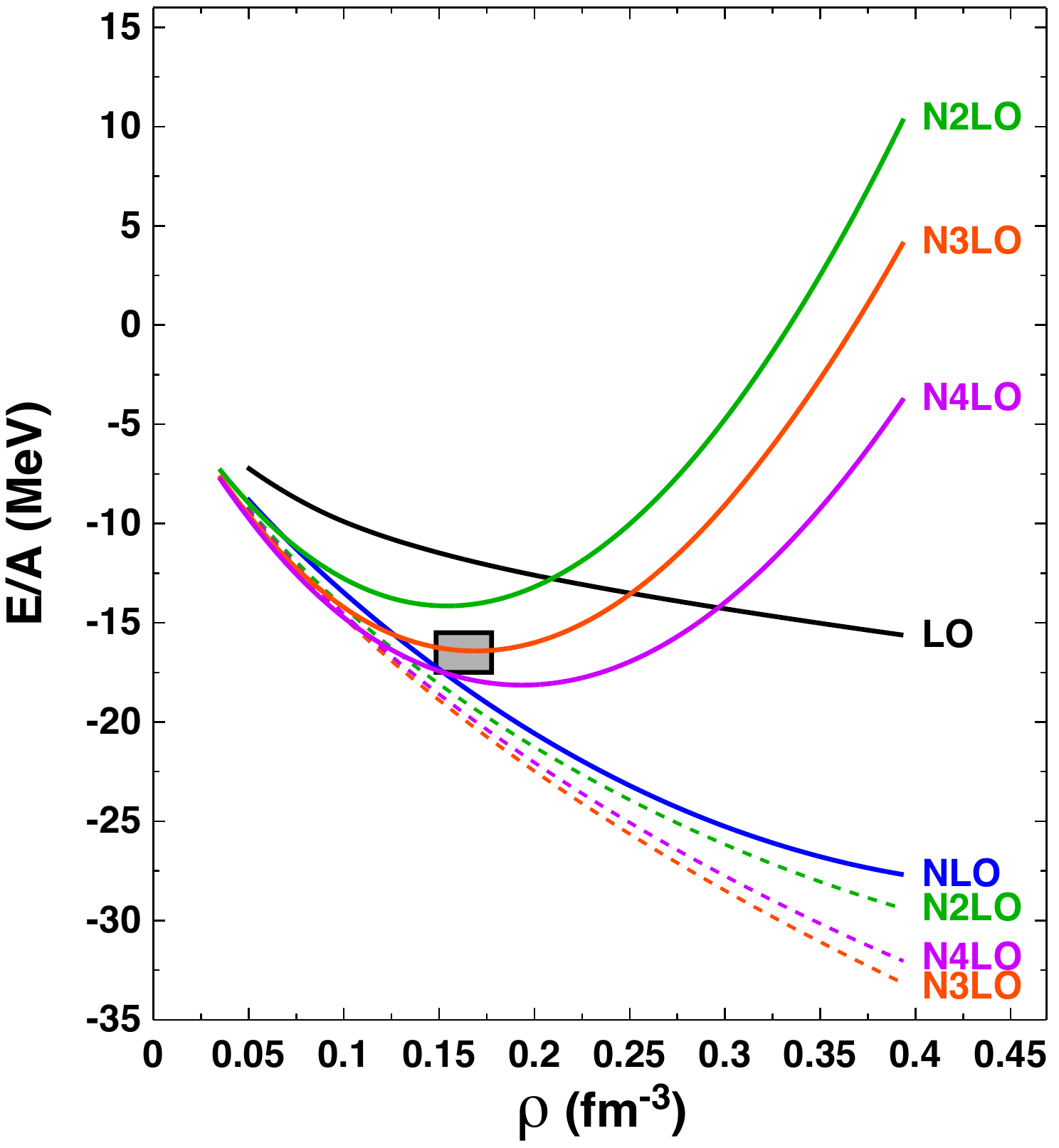}
\caption{Ground state energy per particle of SNM as a function of density from the chiral two-
and three-body forces with cutoff $\Lambda = 450$\,MeV. The three dotted 
curves show predictions which include only two-body forces. For the 3NF contributions at N$^2$LO
and above, the LECs of Table~\ref{tab2} are used. 
The shaded box denotes the approximate empirical saturation energy and density.}
\label{fig2}
\end{figure}

We begin with the study displayed in Fig.~\ref{fig2}, where the momentum-space cutoff is fixed at 
450 MeV but the chiral order of the two-body force is varied from leading to fifth order.
The 3NFs are chosen with LECs in Table~\ref{tab2}, which at N$^3$LO and N$^4$LO include the 
2PE 3NF at fourth and fifth order, respectively.
The dashed lines indicate results at N$^2$LO and above with no three-body forces present, while the 
solid lines include the N$^2$LO three-body force when appropriate.
Formally, we observe a good convergence pattern at the 
two-body level with this family of $NN$ potentials, but naturally we do not expect realistic 
saturation behavior when soft two-body forces alone are
included in the calculation of the EoS.  We see that the inclusion of 3NFs is necessary beyond about
half nuclear matter saturation density and that for this set of nuclear potentials the total 3NF
contribution to the EoS decreases with the chiral order from N$^2$LO to N$^4$LO. 

We note that the uncertainty band obtained by varying the chiral order from N$^2$LO to N$^4$LO while 
keeping $\Lambda = 450$\,MeV fixed encloses the empirical saturation point. 
The saturation energy
varies in the range  $-14\,{\rm MeV} \lesssim E_0 \lesssim -18\,{\rm MeV}$ while the saturation 
density varies between $0.155\,{\rm fm}^{-3} \lesssim \rho_0 \lesssim 0.195\,{\rm fm}^{-3}$. 
We stress that once the two- and three-nucleon forces are fixed by the $NN$ data and the 
properties of the three-nucleon system, no parameters are fined tuned, rendering the many-body 
calculation parameter-free. Since the predicted binding energies and charge radii of 
intermediate-mass nuclei are closely related to the corresponding saturation point in SNM,
we see first evidence that the new class of chiral potentials constructed in this work may lead
to more reliable predictions in ab initio calculations of finite nuclei.
For densities larger than $\rho \gtrsim 0.20\,{\rm fm}^{-3}$, the predictions shown in Fig.~\ref{fig2}
display a trend that does not suggest satisfactory convergence, since the 
three (saturating) solid curves are about equally spaced.
This may be due to the incompleteness of the 3NF at orders above N$^2$LO. 

\begin{figure}[t] 
\centering
\includegraphics[width=7cm]{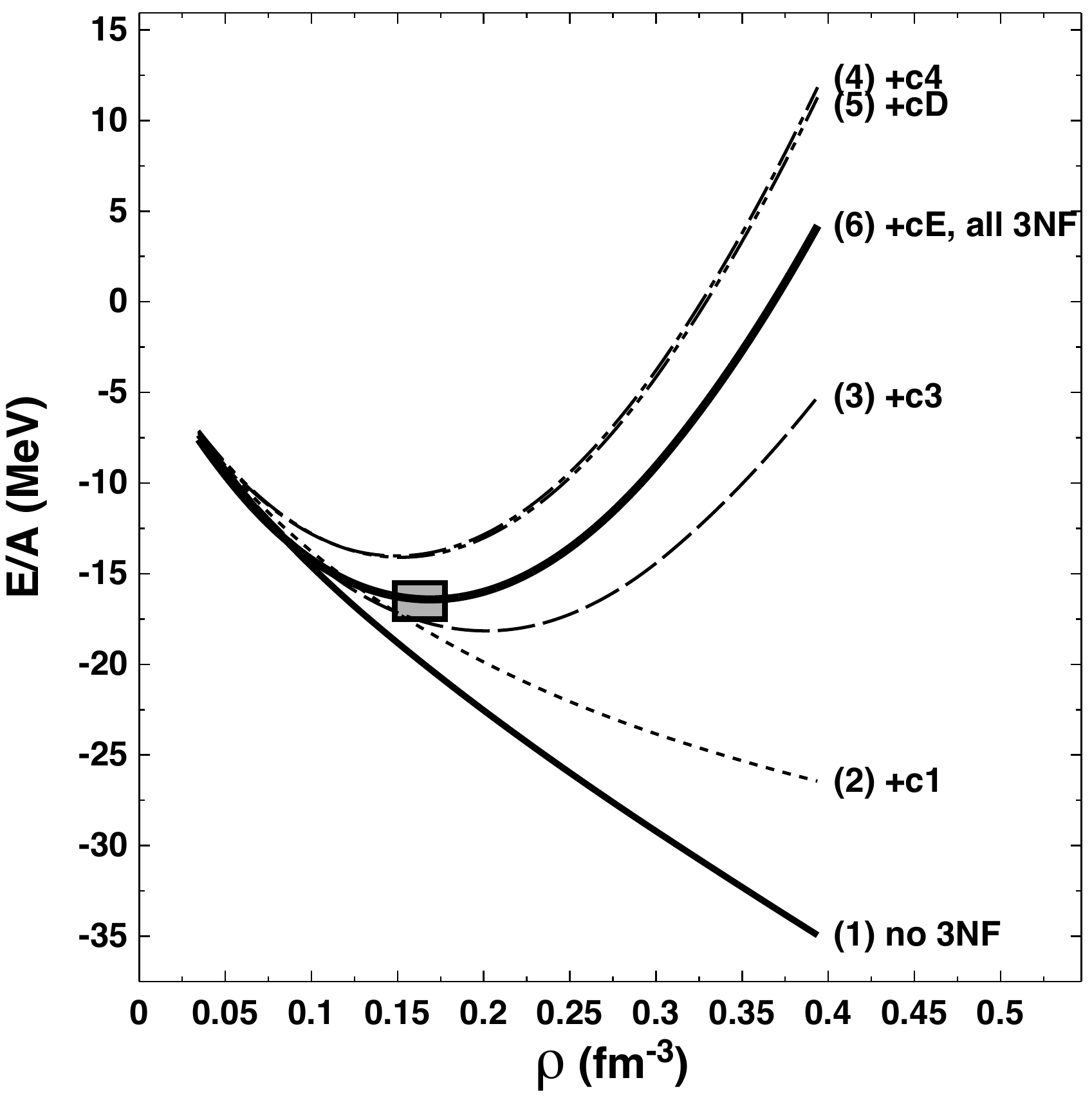}
\caption{Ground state energy per particle of SNM as a function of density from the 
N$^3$LO interaction with a cutoff of 450 MeV. Curves are numbered in ascending order to denote 
that each has been obtained from the previous one with the addition of the indicated 3NF 
contribution. Other details as in Fig.~\ref{fig2}.}
\label{fig3}
\end{figure}

Next we present in Fig.~\ref{fig3} the contribution to the SNM EoS                            
from individual 3NF terms proportional to the LECs $c_{1,3,4}$, $c_D$, and $c_E$.
We have chosen the case of the N$^3$LO $NN$ interaction with a cutoff of 450 MeV together with the
3NF LECs from Table~\ref{tab2}. We observe that the total contribution from terms proportional 
to $c_3$ is very large and instrumental for nuclear saturation. Both the $c_1$ and $c_4$ terms 
generate additional repulsion, while the single contribution from the $c_E$ term is attractive in
this parametrization. Only the $c_D$ term is negligible in this case, even though numerically
it is of natural size.

\begin{figure}[t]
\centering
\includegraphics[width=6.8cm]{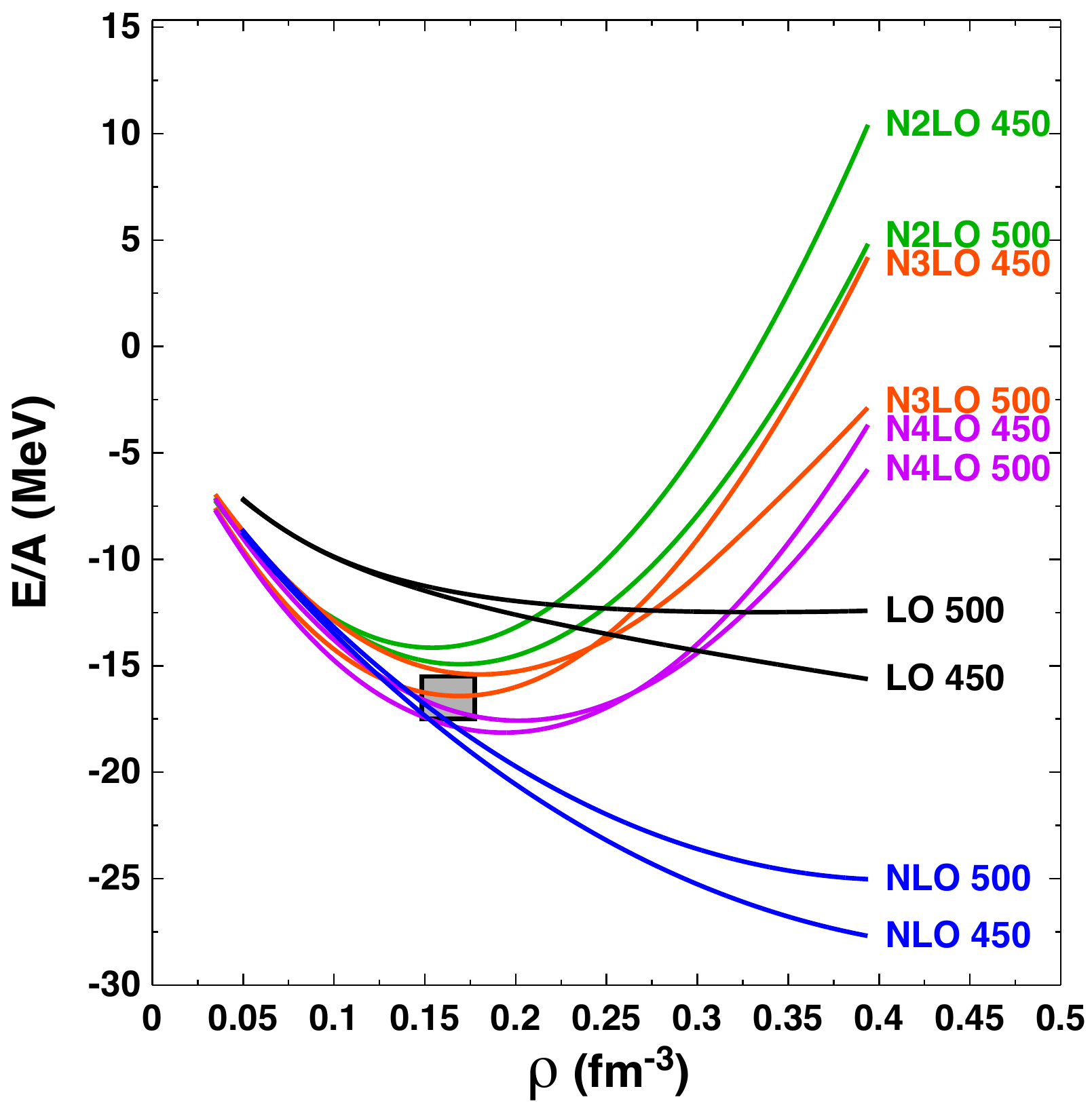}
\caption{Ground state energy per particle of SNM as a function of density at the indicated 
orders and with varying cutoff parameters as denoted. Other details as in Fig.~\ref{fig2}.}
\label{fig4}
\end{figure}

In Fig.~\ref{fig4} we show the dependence of the SNM EoS                                         
on the choice of momentum-space cutoff $\Lambda$ in the two- and three-body forces as
well as the order in the chiral expansion. 
In the present work we consider only the two cases $\Lambda = 450, 500$\,MeV. The nuclear
potentials constructed in \cite{EMN} with $\Lambda = 550$\,MeV were found to have marginal
convergence properties in many-body perturbation theory and therefore not considered in the
present work. At orders N$^2$LO, N$^3$LO, and N$^4$LO, the cutoff dependence appears to be 
comparable but generically smaller than the truncation errors.

\begin{figure}[t] 
\centering
\includegraphics[width=6.85cm]{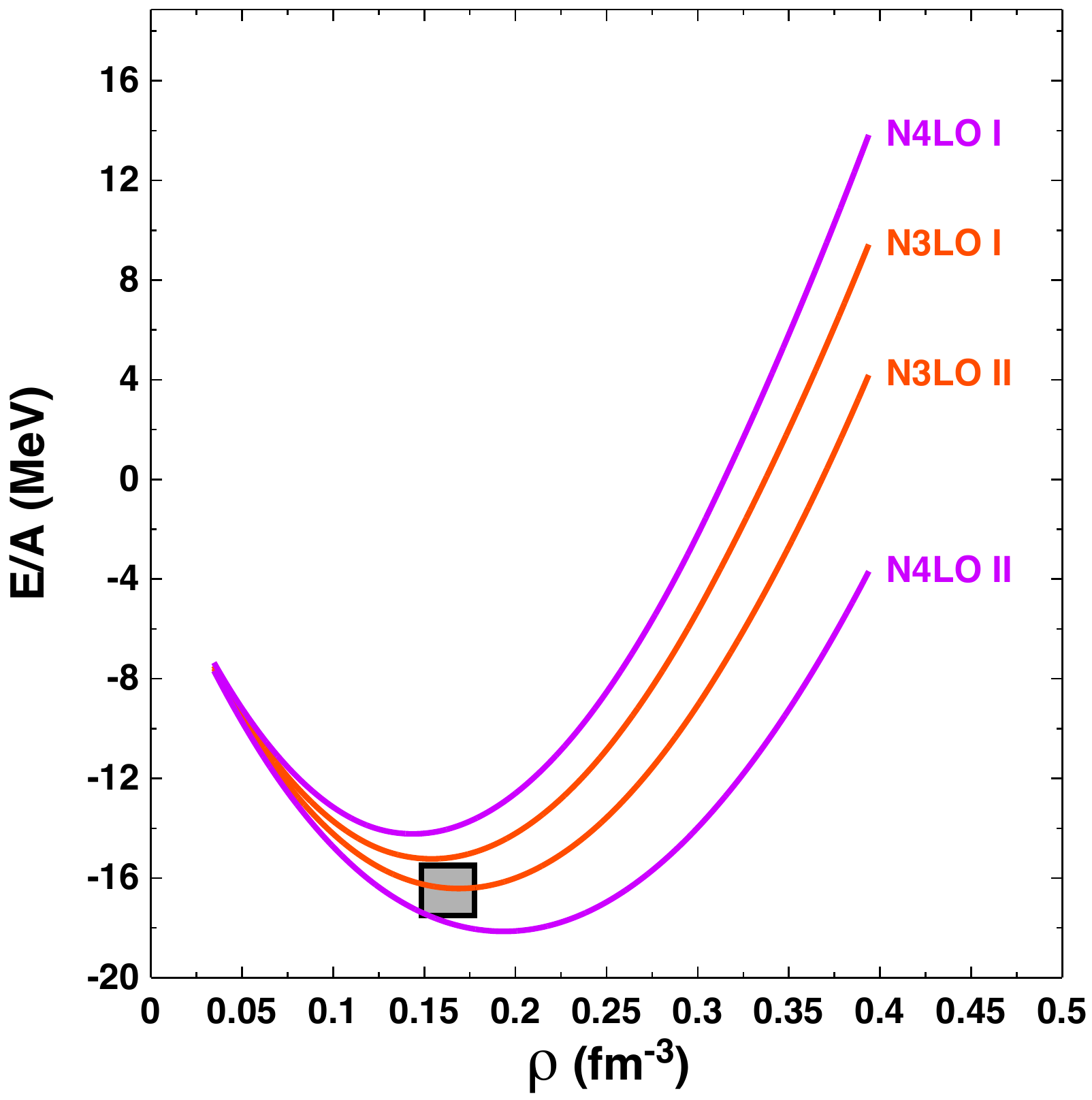}
\caption{Energy per particle in SNM as a function of density at N$^3$LO 
and N$^4$LO with a cutoff of $\Lambda = 450$\,MeV. 
For the 3NF contributions, the LECs of either Table~I or Table~II are applied as indicated by labels `I' or `II', respectively. Case `II' is characterized by including the 2PE 3NF up to the given order.
The shaded box denotes the approximate empirical saturation energy and density.} 
\label{fig5}
\end{figure}

In Fig.~\ref{fig5}, we show the impact of choosing at fourth and fifth order in the chiral
expansion either the 
N$^2$LO 3NF coupling strengths shown in Table \ref{tab1} (labeled ``I'' in the figure) 
or those obtained by including the 2PE 3NF contributions at higher order shown in 
Table \ref{tab2} (labeled ``II'' in the figure). We only show results for potentials with
momentum-space cutoff $\Lambda = 450$\,MeV, but we expect qualitatively similar results
for the $\Lambda = 500$\,MeV cutoff potentials due to the identical change in the important
$c_i$ low-energy constants. We see that at N$^4$LO the impact is rather
large and roughly of the same size as variations in the chiral order from N$^2$LO to N$^4$LO. 
However, the additional theoretical uncertainty resulting from the choice of
LECs entering into the 2PE 3NF would extend only moderately the overall error band inferred 
from Fig.\ \ref{fig4} and only at the largest densities considered.

Finally, we investigate the theoretical uncertainties associated with our many-body method. 
For this purpose we compare the results obtained within the preceding 
nonperturbative particle-particle ladder-diagram summation (NPLDS) technique with calculations 
in many-body perturbation theory (MBPT). This may be especially useful for practitioners 
who intend to employ MBPT for nuclear structure calculations with the new set of chiral forces.
The perturbative expansion of the ground-state energy per particle is   
performed up to third-order in particle-particle ladder diagrams 
within the same formalism found in Ref.\ \cite{Cor13}. In particular, we employ a Hartree-Fock
spectrum for the intermediate-state energies appearing in the second- and third-order diagrams. 
We also calculate the $[2|1]$ Pad\'e
approximant of the MBPT expansion (see Refs.\ \cite{Cor13,Cor14}).
The difference between third-order MBPT in the particle-particle channel, 
$[2|1]$ Pad\'e approximant, and NPLDS gives an 
indication of the quality of the perturbative behavior of the new potentials.

\begin{figure*}[t]
\centering
\includegraphics[scale=0.96]{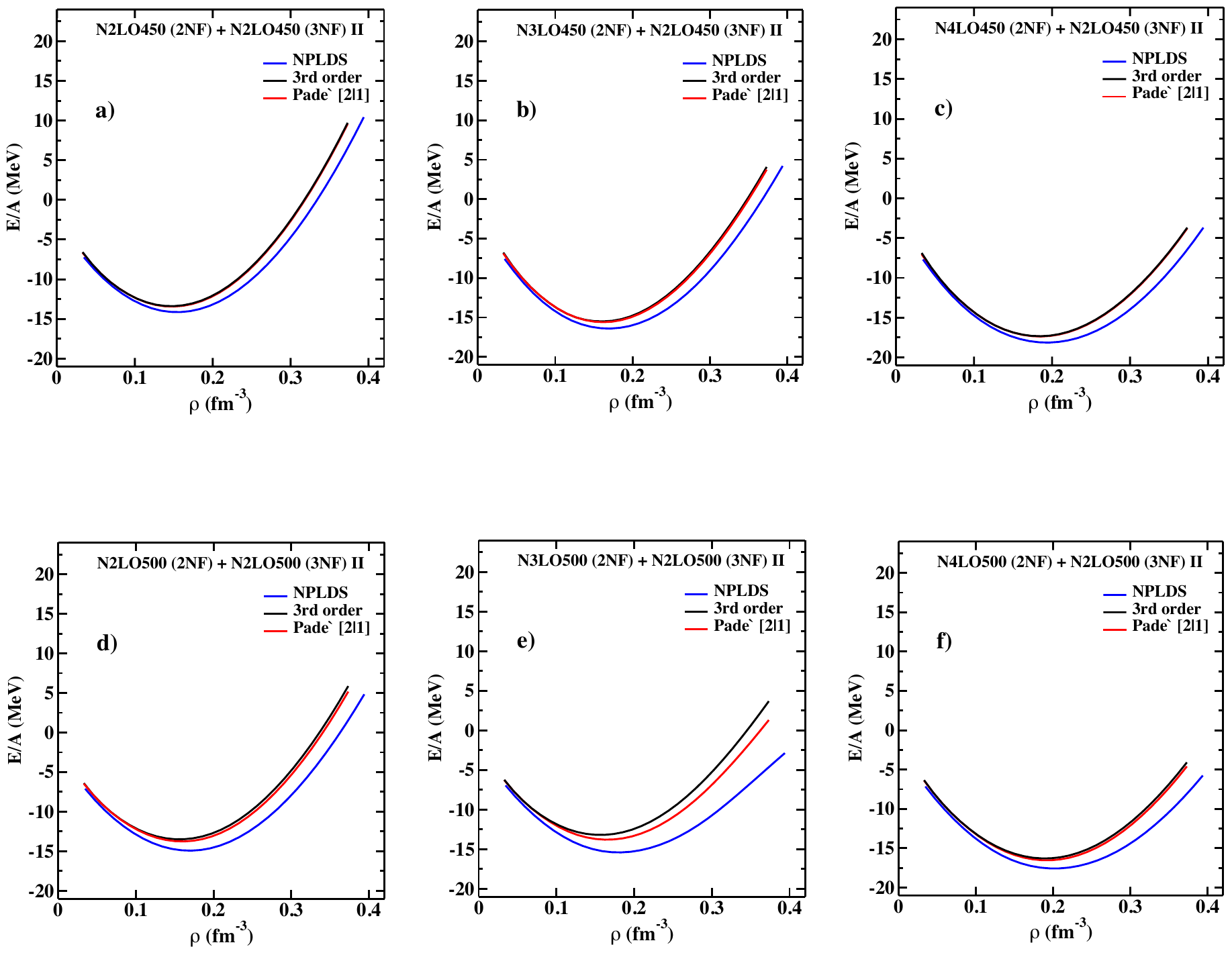}
\caption{The SNM EoS from the $\Lambda = 450$\,MeV (upper panel) and
  the $\Lambda = 500$\,MeV (lower panel) chiral potentials. Results are calculated 
in three ways: (i) NPLDS, (ii) MBPT including up to third-order particle-particle diagrams, and (iii) 
the $[2|1]$ Pad\'e approximant. The 3NF LECs from Table \ref{tab2} are used in all cases.} 
\label{figpert}
\end{figure*}

First, we consider the SNM EoS in the presence of 2NFs only. 
In this case, the results of MBPT are almost identical to those
obtained using either the $[2|1]$ Pad\'e approximant or the NPLDS, the
latter already reported in Figs.\ \ref{fig2} and \ref{fig4}.
We find that the above feature is independent of the order of
the chiral expansion and the choice of the cutoff.
The perturbative nature of the new chiral 2NFs makes 
therefore them suitable for nuclear structure calculations.

On the other hand, inclusion of the 3NF alters the
perturbative behavior of the ground-state energy per nucleon in SNM.
The degree of variation in the perturbative behavior depends on the order in the
chiral EFT expansion and the choice of the cutoff. 
In Fig.~\ref{figpert} , we show the EoS predictions for the $\Lambda = 450, 500$\,MeV 
chiral potentials with 3NF ``II'' low-energy constants
from N$^2$LO to N$^4$LO. Results are obtained for three different scenarios:
third-order MBPT (particle-particle channel), its $[2|1]$ Pad\'e approximant, and the NPLDS.
The differences between the MBPT and NPLDS results are due primarily to the choice
of single-particle spectrum in intermediate-states. In the NPLDS the spectrum is more
compressed relative to the Hartree-Fock spectrum employed in MBPT, leading to an 
enhancement of the large attraction appearing at second-order.
For the larger cutoff value of $\Lambda = 500$\,MeV, the differences between the
NPLDS and MBPT are generically larger. We also observe that the discrepancies between 
the two methods are smaller at N$^2$LO and N$^4$LO as compared to N$^3$LO. 
For $\Lambda = 450$ MeV, however, the perturbative behavior is quite good across all
scenarios considered.

We conclude this section with additional comments on the important issue of SNM saturation. 
In the recent work of Ref.~\cite{Sim17}, the aim was to explore the impact of SNM saturation on 
the properties of closed and open-shell nuclei with $A \leq$ 78. 
In another investigation~\cite{Mor17}, in which the same forces were applied,
the structure of the light Sn isotopes were studied, reproducing both the binding energy and the
small splitting between the lowest $J^\pi=7/2^+$ and $5/2^+$ states of $^{100}$Sn.
In those studies, the N$^3$LO potential ($\Lambda = 500$\,MeV) of Ref.~\cite{EM03} was taken as the 
starting point and then evolved through similarity renormalization group (SRG) transformation into 
low-resolution interactions (i.e., soft interactions). These SRG-evolved $NN$ interactions were 
then combined with the leading chiral 3NF (with $c_D$ and $c_E$ coupling strengths refitted to
the $^3$H binding energy and $^4$He charge radius) to 
obtain various Hamiltonians with different cutoff combinations~\cite{Heb11}. 
Ideally, it would be preferable to construct from the outset perturbative two-body forces that do not
require additional RG evolution with associated uncontrolled induced many-body forces. 
From the results of this section, we suggest that the new set of chiral nuclear forces
developed in the present work may produce equally favorable results as those of
Refs.~\cite{Heb11,Sim17,Mor17}, and we encourage studies of nuclear systematics based upon 
these potentials. 

\subsection{Neutron matter: predictions and discussion} 
\label{IVb} 

We next consider the ground state energy of pure neutron matter as a function of density,
employing the same set of chiral
potentials and many-body methods discussed previously in the case of symmetric nuclear matter. 
The EoS for both SNM and PNM are crucial to determine the density-dependent nuclear symmetry 
energy and to better understand the properties of neutron-rich nuclei and neutron stars.

\begin{figure}[t] 
\centering
\includegraphics[width=7.cm]{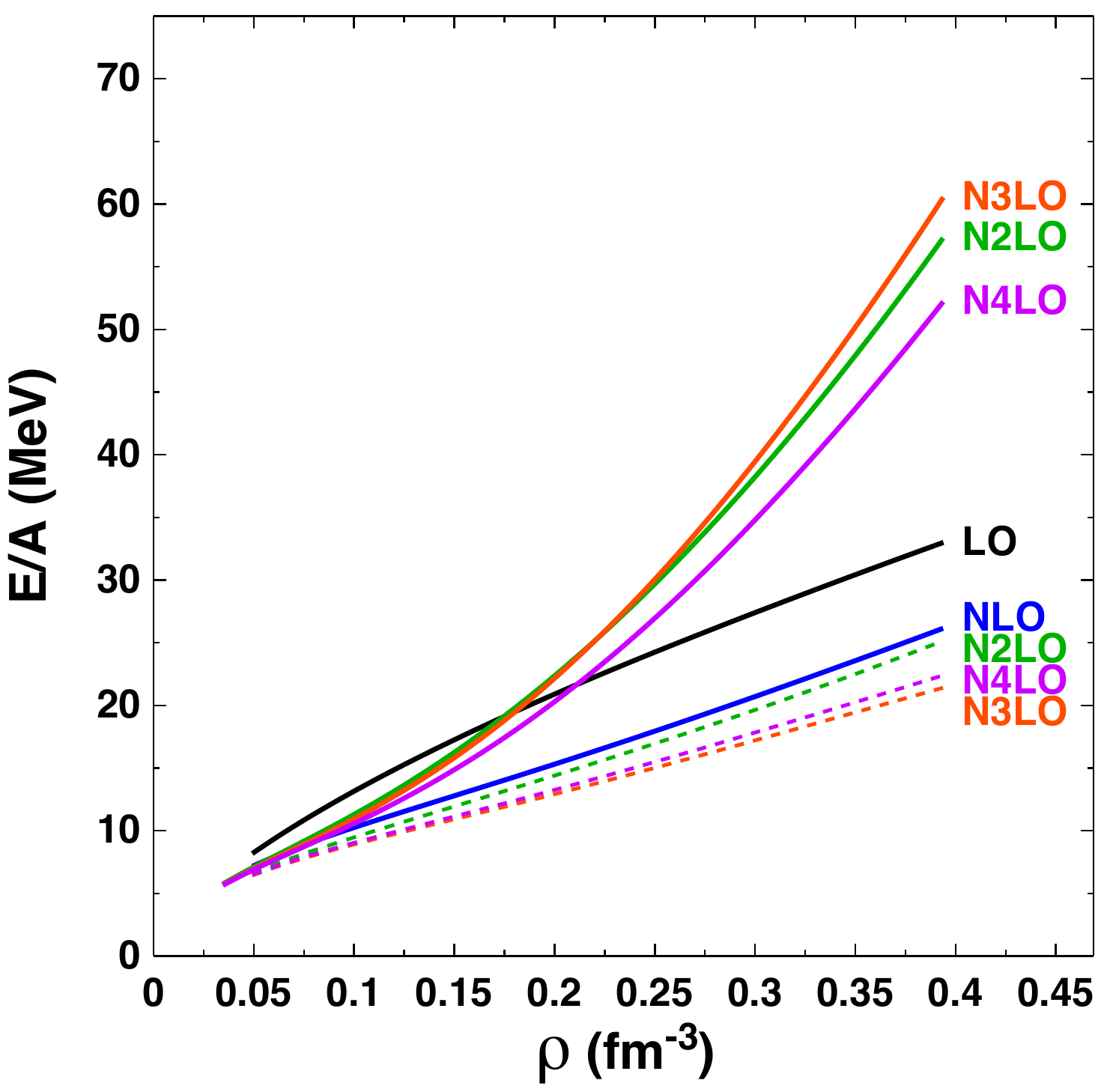}
\caption{Ground state energy per particle of PNM as a function of density at the indicated orders
in the chiral expansion. 
The three dotted curves show predictions including only the 2NF. The cutoff parameter is fixed at 
$\Lambda = 450$\,MeV and the 3NF LECs from Table~\ref{tab2} are used.}
\label{fig6}
\end{figure}

\begin{figure}[t] 
\centering
\includegraphics[width=8.05cm]{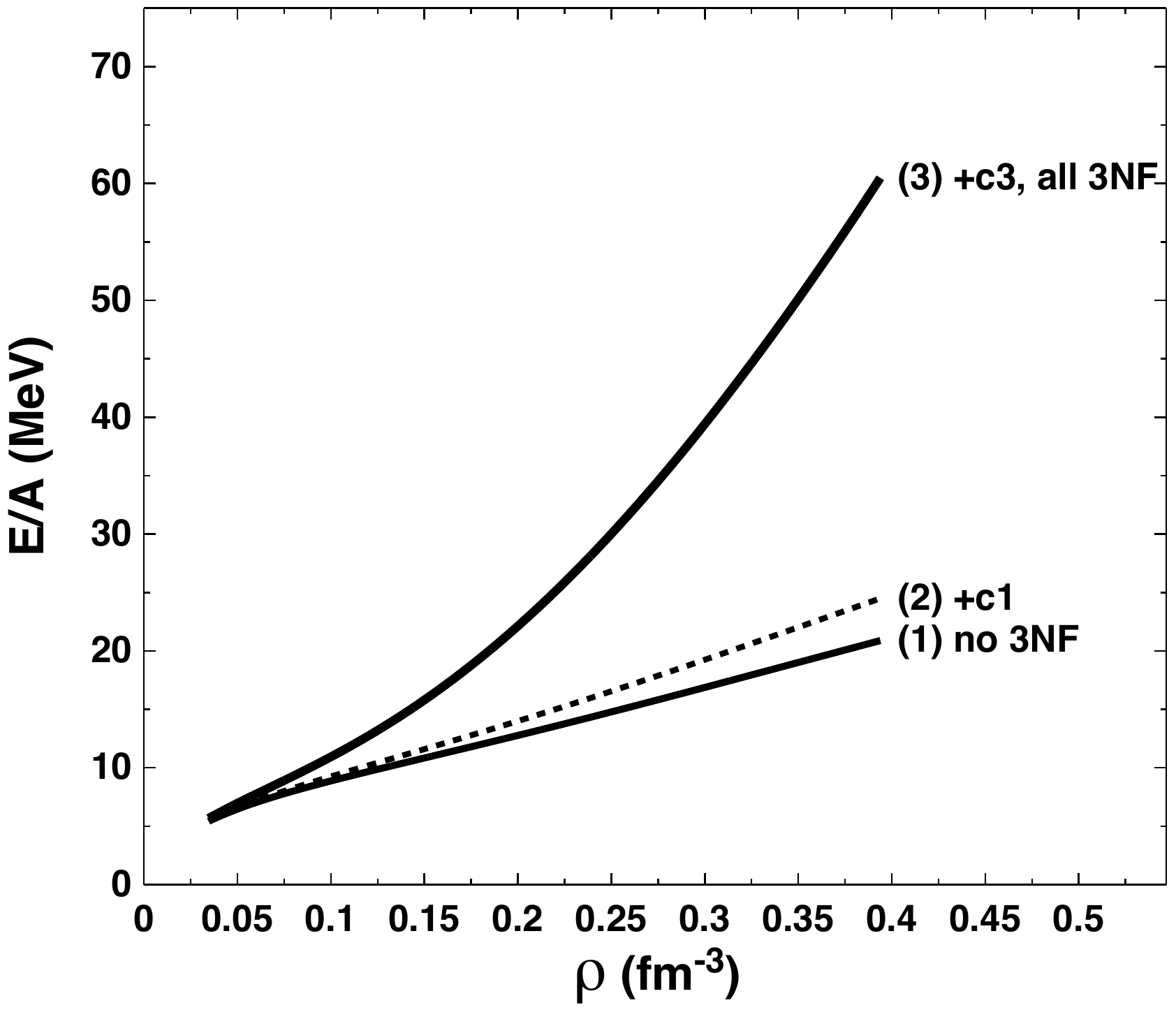}
\caption{Ground state energy per particle in PNM as a function of density at N$^3$LO and 
with a cutoff of 450 MeV. Curves are numbered in ascending order to signify that each has 
been obtained from the previous one with the addition of the indicated 3NF contribution. 
Note that the 3NF contributions depending on  $c_4$, $c_D$, and $c_E$ vanish in PNM.
The LECs of Table~\ref{tab2} are used. } 
\label{fig7}
\end{figure}

In Fig.~\ref{fig6} we show the energy per particle of PNM as a function of density starting from 
chiral two- and three-body forces with the same value of the momentum-space cutoff
$\Lambda = 450$\,MeV but at different orders in the chiral expansion. As in the case of symmetric 
nuclear matter, we observe good convergence at the level of 2NF alone.
When 3NFs are included, we find somewhat smaller truncation errors compared to the case of 
SNM. This may be due in part to the absence of large, central isospin-0 partial waves in PNM, 
which appear to be more sensitive to differences among interactions. As shown in Fig.~\ref{fig7},
the low-energy constant $c_3$ is responsible for essentially all of the additional repulsion generated
from the two-pion-exchange three-body force. Clearly, the 3NF 
plays an outstanding role in very neutron-rich systems at and beyond nuclear saturation density, where
its contribution to the EoS grows more strongly with the density than the 2NF contributions. 

\begin{figure}[t] 
\centering
\includegraphics[width=7.06cm]{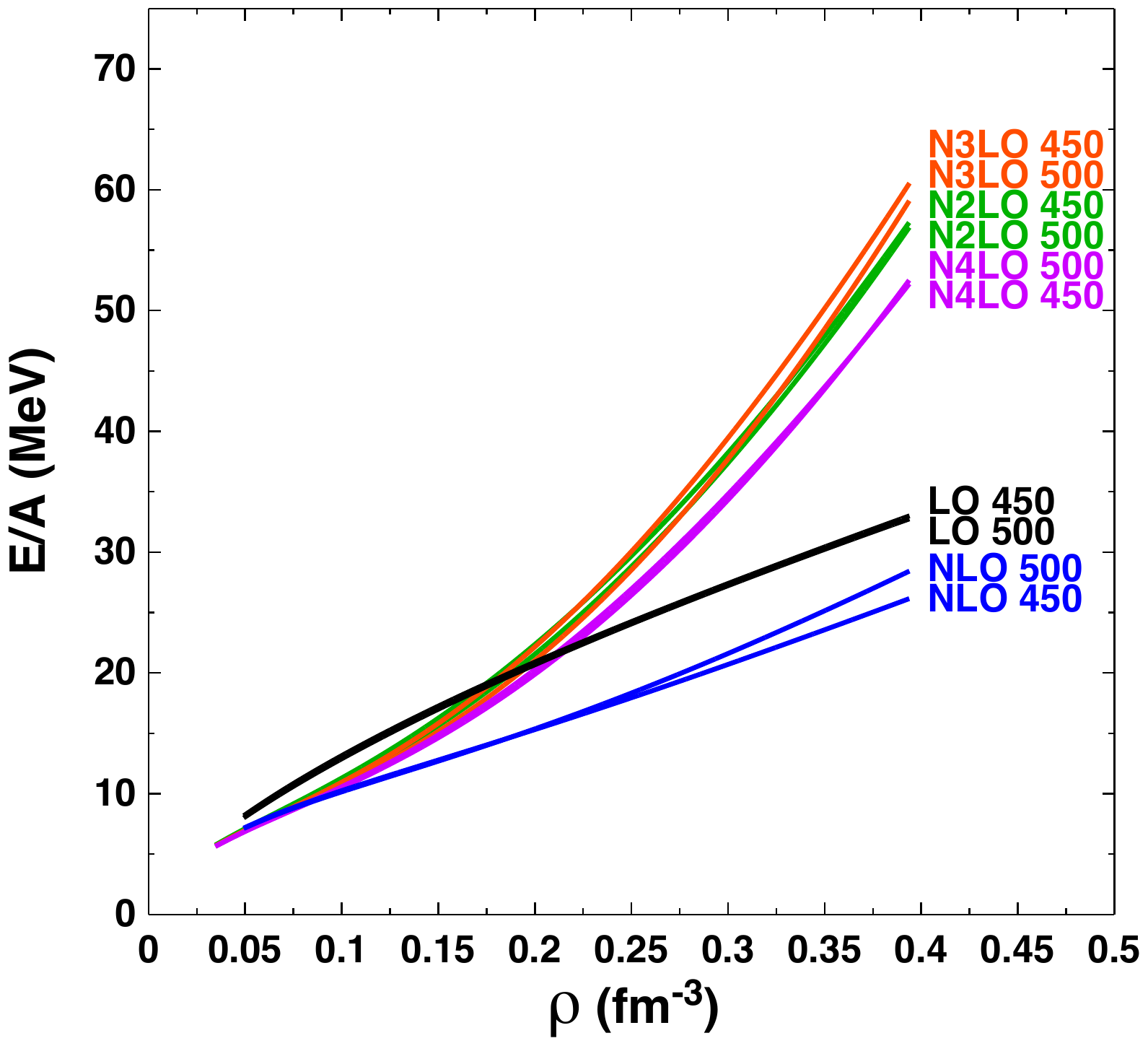}
\caption{Ground state energy per particle of PNM as a function of density at the indicated chiral 
orders and with varying cutoff parameters as denoted. The LECs of Table~\ref{tab2} are used.}    
\label{fig8}
\end{figure}

We show in Fig.~\ref{fig8} the energy per particle of pure neutron matter as a function of density when
varying both the order in the chiral expansion and the momentum-space cutoff 
$\Lambda = 450, 500$\,MeV. We see that in comparison to the analogous study in symmetric nuclear
matter, the pure neutron matter results display a
much weaker cutoff dependence, which may again be due to the absence of strong isospin-0 partial
waves. Interestingly, even in the case of the relatively large density $\rho = 0.4$\,fm$^{-3}$
corresponding to a Fermi momentum of $k_F = 450$\,MeV that lies at the effective breakdown 
scale of the expansion, there is relatively little cutoff dependence.

\begin{figure}[t]
\centering
\includegraphics[width=7.1cm]{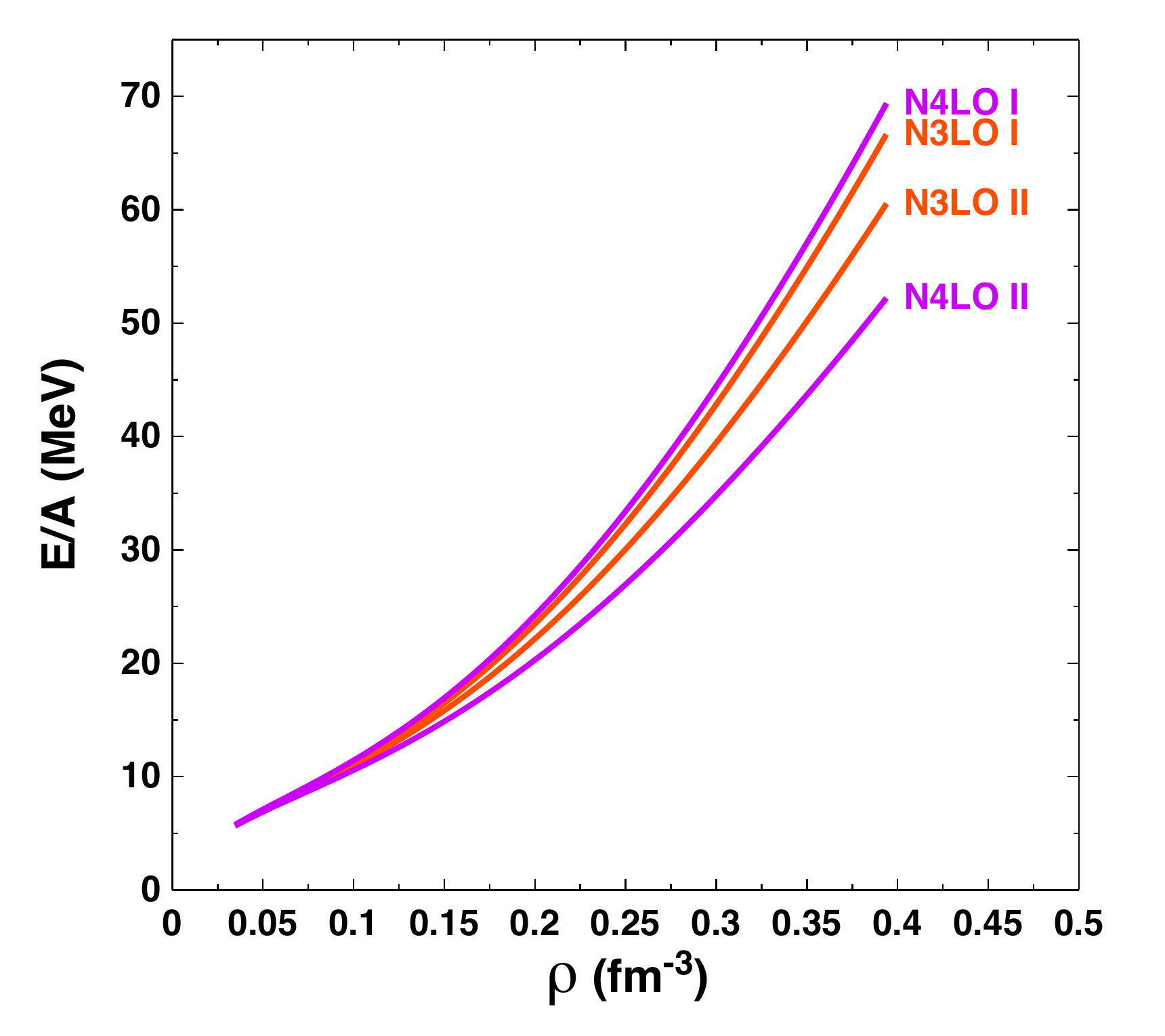}
\caption{Ground state energy per particle of PNM as a function of density at N$^3$LO 
and N$^4$LO with a cutoff of 450 MeV. Similar to Fig.~\ref{fig5},
for the 3NF contributions the LECs of either Table~I or Table~II are applied as indicated by labels `I' or `II', respectively.}
\label{fig9}
\end{figure}

The impact of including the 2PE 3NF at fourth and fifth order, compared to including only the third-order
contributions, through the adoption of the LECs given in Table~\ref{tab2} is demonstrated in Fig.~\ref{fig9}. 
As in the case of symmetric nuclear matter, the effect at N$^4$LO is much larger than that at N$^3$LO due 
to the larger change $\Delta c_3 = 1.94$\,GeV$^{-1}$ vs.\ $\Delta c_3 = 0.89$\,GeV$^{-1}$, respectively,
in the $c_3$ low-energy constant at these two orders in the chiral expansion.
Moreover, the choice of LECs entering into the 2PE 3NF contributions again results in a moderate
systematic increase in the pure neutron matter energy per particle at the highest densities 
considered. It will be interesting to explore the influence of such higher-order 3NF contributions
on predictions of neutron star properties. This investigation is in progress.

Regarding the perturbative properties of the PNM EoS, 
we have found that the results obtained using MBPT up to third
order in the particle-particle channel, with and without inclusion of 3NF, 
are consistent with those shown in Figs. \ref{fig6} -- \ref{fig9}.
The good perturbative behavior is independent of the choice of
cutoff $\Lambda$ as well as the chiral order at which      
the nuclear potentials have been derived. We also find the results to be 
consistent with those obtained in Ref.\ \cite{Cor13}.                                    

\section{Conclusions and outlook}                                                                  
\label{Concl} 

In the present work we have extended the new class of chiral $NN$ interactions from LO 
to N$^4$LO in Ref.\ \cite{EMN} to include the complete set of
three-nucleon forces at N$^2$LO in the chiral
expansion. The 
additional LECs $c_D$ and $c_E$ that appear at this order are fitted to the binding 
energies of $A=3$ nuclei and the $\beta$-decay lifetime of $^3$H. We have included as well the 
two-pion-exchange three-body forces at N$^3$LO and N$^4$LO in order to probe the effect of missing 
higher-order many-body contributions. The resulting set of nuclear potentials were 
then employed in studies of the nuclear and neutron matter EoS~\cite{note}.
We find quite good nuclear matter saturation properties together
with relatively small uncertainties in the PNM EoS even up to relatively large
densities. Moreover, the calculation of the EoS within many-body
perturbation theory exhibits quite good behavior with cutoff $\Lambda=450$\,MeV and satisfactory
behavior with cutoff $\Lambda=500$\,MeV.
While complete calculations at N$^3$LO and N$^4$LO are the ultimate goal, we suggest 
that the present set of high-quality potentials will
be valuable for systematic studies of medium-mass and heavy nuclei as well as studies of
neutron star structure.

\section*{Acknowledgments}
The work of F.S.\ and R.M.\ was supported in part by the U.S.\ Department of Energy, Office of Science, Office of Basic Energy Sciences, under Award Number DE-FG02-03ER41270. The work of J.W.H.\ was supported by the National Science Foundation under Grant No.\ PHY1652199 and the NNSA through DOE Cooperative Agreement DE-NA0003841. 
Computational resources provided by the INFN-Pisa Computer Center are gratefully acknowledged.


\begin{references}
\bibitem{note}
A similar investigation has been conducted in Ref.~\cite{DHS17}, where 
a different philosophy for the adjustment of the 3NF is pursued and a
complete N$^3$LO 3NF is applied.
\end{references}
\end{document}